\documentstyle[prc,psfig,eqsecnum,aps]{revtex}

\begin{document}
\draft
\title{ The realistic collective nuclear Hamiltonian}
\author{Marianne Dufour and Andr\'es Zuker}

\address{Physique Th\'eorique, B\^at 40/1 CRN,
  IN2P3-CNRS/Universit\'e Louis Pasteur BP 28, F-67037 Strasbourg
  Cedex 2, France}
\date{\today}

\maketitle

\begin{abstract}
  The residual part of the realistic forces ---obtained after
  extracting the monopole terms responsible for bulk properties--- is
  strongly dominated by pairing and quadrupole interactions, with
  important $\sigma\tau\cdot\sigma \tau$, octupole and hexadecapole
  contributions. Their forms differ from the schematic ones through
  normalizations that lead to a universal $A^{-1/3}$ scaling for all
  multipoles. Coupling strengths and effective charges are calculated
  and shown to agree with the observed values.
\end{abstract}


In the preceeding paper \cite{ZD} the nuclear Hamiltonian was shown to
separate into an ``unperturbed'' monopole field ${\cal H}_m$ - that
demands phenomenological treatment - and a
``residual'' multipole contribution, ${\cal H}_M$, that is well given by
existing realistic interactions \cite{ACZ91,CPZM94}. In principle the
task of fully characterizing ${\cal H}$ becomes a pure monopole one,
but for this characterization to be of practical use it is necessary to
understand the structure of  ${\cal H}_M$, which is our present task.

It is through empirical evidence that we know that it must be possible
to describe nuclei by solving the Schr\"odinger equation: the systems
appear to be rather dilute - in the sense that the residual interaction
is weak enough to be treated in perturbation theory - except for few
(sub)shells in the vicinity of the Fermi level.

The quality of the interactions can be tested
convincingly only in cases where exact diagonalizations are feasible in spaces
large enough (typically one major shell) to ensure that what is left out
 can be treated perturbatively. The number of possible tests - though
 limited to vector spaces that do not exceed dimensionalities $O(10^7)$ -
 is significant, and it is certainly no accident that the realistic
 ${\cal H}_M$ passes them all rather well. Therefore, the time may have come
 to {\em trust} the realistic forces rather than {\em test} them.

 There are many advantages in doing so, but one is special.
   As new regions open to scrutiny - through Monte Carlo techniques \cite{ALHA}
  and further improvements in shell model technology \cite{NOWA} - the nature
  of the problems  changes: Dimensionalities grow
  exponentially with the size of the systems, but the behaviour described by
  the enormous matrices also becomes simpler through the increasing influence
  of coherent effects.

The special advantage in trusting the Hamiltonian is that
in adapting existing methods to face this situation, or designing new ones,
our basic tool is the Hamiltonian itself.

It is here that a deeper understanding of ${\cal H}_M$ is called for.

\medskip

In section I we are going to show  that its structure is as simple as could
be expected, because of the strong dominance of  pairing and lowest
multipole operators. They appear in a {\bf normalized} form that retains
 all the simplicity of their traditional
counterparts but suffers none of their drawbacks.

In section II, it is explained how to ensure that a given choice of
dominant terms is optimally extracted from the Hamiltonian.

Section III presents some estimates of the renormalizations due to core
polarization effects.

Section IV is devoted to a comparison of different realistic interactions
among themselves, and with the $W$ fit in the sd shell  \cite{BW88}.

Appendix A contains results on the matrix defining the diagonal multipole
representation.

Appendix B revisits the pairing plus quadrupole model.

\section{Diagonal Representations of ${\cal H}$}
Our plan is as follows: a) Introduce  ${\cal H}_M$ in the two standard
representations, b) reduce to sums of separable terms by diagonalization,
c) show thay the few dominant ones are {\it normalized} versions of the
standard pairing and multipole forces, d) show that the normalizations are
dictated by the universal $A^{-1/3} $  scaling for all couplings; and finally
in e) say a few words about monopole the consequences of this result.

\smallskip

{ \bf {a) The Hamiltonian}}. We start by borrowing eqs. (I.41 to I.43)
of \cite{ZD} to write  ${\cal H}_M$ in the normal and mutipole representations:

  \begin{mathletters}
  \begin{equation}
{\cal H}_M=\sum\limits_{r\leq s,t\leq u,\Gamma}W_{rstu}^\Gamma Z_{rs\Gamma}^+
\cdot Z_{tu\Gamma},\label{1}
\end{equation}
\begin{equation}
{\cal H}_M=\sum_{rstu \Gamma}[\gamma]^{1/2}{(1+\delta_{rs})^{1/2}
(1+\delta_{tu})^{1/2}\over4}\omega_{rtsu}^\gamma (S_{rt}^\gamma
S_{su}^\gamma)^0,\label{2}
\end{equation}
\begin{equation}
\omega_{rtsu}^\gamma =\sum_\Gamma(-)^{s+t-\gamma-\Gamma}
\left\{\begin{array}{ccc}
   r&s&\Gamma\\  u&t&\gamma
   \end{array}\right\}
W_{rstu}^\Gamma[\Gamma],\label{3a}
\end{equation}
\begin{equation}
W_{rstu}^\Gamma =\sum_\gamma(-)^{s+t-\gamma-\Gamma}
\left\{\begin{array}{ccc}
   r&s&\Gamma\\  u&t&\gamma
   \end{array}\right\}
\omega_{rtsu}^\gamma[\gamma],\label{3b}
\end{equation}
\end{mathletters}
 Since ${\cal H}_m$ is defined as containing all the $\gamma=00$ and 01
 terms, ${\cal H}_M$ is defined by
$$\omega_{rstu}^\gamma=0\quad{\rm  for}\quad\gamma=00\quad{\rm  and}\quad 01$$
which explains the absence of one body contractions.

The normal or $W$-form is unique and the ordering of the indeces simply
eliminates double counting: the contributions in $rtsu$,$rstu$,$rsut$ and
$rust$ are identical, and it is just as well to keep only the first.

The multipole or $\omega$-form is highly non unique because the terms are
not linearly independent and permuting indeces leads to different objects.
We have chosen the variant in which summations are unrestricted for a reason
that will become immediately apparent.

For the calculations we adopt the KLS force \cite{KLS69,LEE69},
but in section IV the claim that the choice of {\it realistic} interaction
does not matter much.

\smallskip

{\bf{b) Separable form}}. We call $H_M$ the restriction of ${\cal H}_M$ to
a finite set of orbits.
Replacing pairs by single indeces $rs\equiv x$, $tu=y$ in (1) and
$rt\equiv a$, $su=b$ in (2), and bringing the matrices
$W_{xy}^\Gamma$ and $f_{ab}^\gamma=\omega_{ab}^\gamma
\sqrt{(1+\delta_{rs})(1+\delta_{tu})}/4$, to diagonal form through
unitary transformations $U_{xk}^\Gamma,u_{ak}^\gamma$:
\begin{mathletters}
\begin{equation}
U^{-1}WU=E\longrightarrow W_{xy}^\Gamma =\sum_k
U_{xk}^\Gamma U_{yk}^\Gamma E_k^\Gamma \label{III.3}
\end{equation}
\begin{equation}
u^{-1}fu=e\longrightarrow f_{ab}^\gamma =\sum_k
u_{ak}^\gamma u_{bk}^\gamma e_k^\gamma, \label{III.4}
\end{equation}
\end{mathletters}
and then,
\begin{mathletters}
\begin{equation}
H_M=\sum_{k,\Gamma}E_k^\Gamma\sum_xU_{xk}^\Gamma
Z_{x\Gamma}\cdot\sum\limits_yU_{yk}^\Gamma Z_{y\Gamma},\label{4a}
\end{equation}
\begin{equation}
H_M=\sum_{k,\gamma}e_k^\gamma\left(\sum_au_{ak}^\gamma
S_a^\gamma\sum_bu_{bk}^\gamma S_b^\gamma\right)^0
[\gamma]^{1/2},\label{4b}
\end{equation}
\end{mathletters}
which we call the $E$ and $e$ representations. Note here the explanation of
the unrestricted ordering of the orbital indeces: it guarantees that in the
$f_{ab}$ matrices, $a$ and $b$ belong to the same set. In  Appendix A it
is explained what happens when they do not ({\it asymmetric factorization}).

\smallskip

{\bf{c) Dominant terms}}.
We have calculated the eigensolutions in (4) using KLS for spaces of one and
two major oscillator shells. The density of eigenvalues (their number in
a given interval) in the $E$
representation is
shown in fig.~1 for a typical two-shell case.
It is skewed, with a tail at negative
energies which is what  we expect from an attractive interaction.

The $e$ eigenvalues have a number of simple properties demonstrated
in  Appendix A: their mean value always vanishes, their width is $\sqrt{1/8}$
 of that of the $E$ distibution, and they are twice as numerous.
  In fig.~2 we find that they are very symmetrically distributed around a
narrow central group, but few of them are neatly detached. The strongest
 have $\gamma^\pi=1^-0,\;1^+1,~2^+0,~3^-0, 4^+0$. {\it If the
corresponding states are eliminated from H in} (4b) {\it and the associated
H in} (4a) {\it is recalculated, the E distribution becomes quite
symmetric}. Details will be given in section II, and here we only note
that the residual skewness is entirely accounted for by the $\Gamma=1^+0,\;
0^+1
\;and\; 2^+0$ peaks, of which the first remains strong at  $-7$MeV.

This result is most telling because from the work of Mon and French \cite{MF75}
 we know that a symmetric $E$ distribution will lead to spectra in the
 $n$-particle systems that are identical to those of a random matrix.
 Therefore, we have found  that - with the
 exception of three $\Gamma$ peaks - the very few dominant states in the
 $e$-distribution are responsible for deviations from random behaviour in
$H_M$.
 Positively stated, these states are at the origin of collective properties.

If the diagonalizations are restricted to one major shell, negative parity
peaks are absent, but for the positive parity ones the results are
practically identical to those of figs.~1 and 2, except that the energies
are halved, a striking feature whose significance will become clear soon.

  In the list of important contributions whose structure we analyze
 we include the $\Gamma=10$ and 01 terms, and the six strongest $\gamma$ ones

 Their eigenstates (i.e. the factors in eqs.(4) with $k=1$) will be compared
with standard pairing and multipole operators. To fix ideas we write the
form these eigenstates should take in the case of perfect pairing ($\Gamma=01$)
and quadrupole forces ($\gamma=20$) acting in one shell of principal quantum
number $p$. To compare with the result of a diagonalization, the operators
must be  normalized:
\begin{eqnarray}
\overline P\,^+_p\equiv \overline P\,^+_{01p}=
\sum\limits_{r\in p}Z_{rr01}^+\Omega_r^{1/2}/\Omega_p^{1/2},\;
\Omega_p\cong 0.655\, \rm A_{mp}^{2/3}\label{5a}\\
\bar q_p\equiv M_p^{20}=
\sum\limits_{rs\in p}S_{rs}^{20}q_{rs}/{\cal N}_p,\quad
{\cal N}_p^2\cong 0.085\, \rm A_{mp}^{4/3}\label{5b}
\end{eqnarray}
where
\begin{itemize}
\item $\Omega_r=j_r+1/2$, $q_{rs}=<r\| r^2Y^2\| s>/\sqrt 5$
\item $A_mp$ is the the total number of particles at midshell $p$
($p^{(2)}=p(p-1)$, remember)
\begin{eqnarray*}
 A_{mp}=2\sum_{p'< p}(p'+2)^{(2)}+(p+2)^{(2)}=\\
{1\over 3}
((p+1)(p+2)(2p+3)\approx{2\over 3}(p+3/2)^3.
\end{eqnarray*}
\item
\ the norms $\Omega_p$ and  ${\cal N}_p$, are then

\begin{eqnarray*}
\Omega_p=\sum_{r}\Omega_r={(p+2)^{(2)}/ 2}\approx
{1\over 2}\left({{3\rm A_{mp}}/ 2}\right)^{2/3},\\
{\cal N}_p^2=\Sigma q_{rs}^2\cong 5
(p+3/2)^4/32\pi=\frac{5}{32\pi} (3A_{mp}/2)^{4/3}.
\end{eqnarray*}
For the calculation of ${\cal N}_p^2$ we have used the
matrix elements listed in (A11-15) of [1].
\end{itemize}

For the other strong multipoles the choice of operators is evident
and for $\Gamma=10$ the simple idea is that pairing in LS coupling
should produce a good candidate. Labelling the orbits by their $\ell$
quantum numbers, we have two pairing terms
\begin{eqnarray}
 \bar P_{ST}^+=(\sum_\ell[ \ell])^{-1/2}\sum_\ell[\ell]^{1/2}
Z_{\ell\ell0ST}^+ \quad ST=01,10,\nonumber\\
\text {which in } jj \text { coupling become }
\bar P_{01}^+=\Omega_p^{-1/2}\sum_r \sqrt{\Omega_r} Z_{rr01}^+
\nonumber\\
\bar P_{10}^+=\Omega_p^{-1/2}\sum_{j,j'}[\ell]^{1/2}\{(\ell\ell)
0\,({1/2}\,{1/2})1\,| (jj')1\}  Z_{rr'10}^+\,,
\end{eqnarray}
where we have recovered the usual $r\equiv j\ell$ label,and used a
self evident notation for the LS to $jj$ transformation.
In table~I these operators are compared with the results of the
diagonalization.
It is apparent that $ \bar P_{p01}$ accounts very well for $(U_{rr}^{01})_p$.
For $ \bar P_{p10}$ {\it vs} $(U_{rs}^{10})_p$
the agrement is not so excellent, but still good. The  overlaps are found
under $<U_p| P>$) in table~II, which also contains the corresponding
values of $<u_p| M>$ for the  lowest multipole operators $M$. The agreement
is again excellent except for the $\sigma$ case, for which it is only fair.
 Note that the form of these operators is given in the Appendix of [1].
\begin {table}
\caption{Eigenvectors and energies calculated in the $pf(p=3)$, $sdg(p=4)$, and
$pf+sdg(3+4)$ spaces compared with the normalized pairing operators
$\bar P_{10}$ and $\bar P_{01}$}
\begin {tabular}{|c|c|c|c|c|c|c|} \tableline
& $\bar P_{01}$&$(U_{rs}^{ 01})_{3,4}$&$\sqrt{2}(U_{rs}^{01})_{3+4}$&
$\bar P_{10}$&$(U_{rs}^{10})_{3,4}$&$\sqrt{2}(U_{rs}^{10})_{(3+4)}$\\
\tableline
 77&-.63&-.65&-.66&.41&.31&.31\\
 75&&&&-.68&-.74&-.74\\
 33&-.45&-.38&-.41&.33&.26&.27\\
 35&&&&0.&-.30&-.25\\
 31&&&&-.42&-.43&-.44\\
 55&-.55&-.58&-.59&-.27&-.12&-.13\\
 11&-.32&-.29&-.31&-.04&.07&.04\\
\tableline
 $E_{\Gamma(3)}$&&-2.95&& &-4.59&\\
 \tableline
 99& .58& .61& .64&-.37&-.26&-.25\\
 97&    &    &    & .63& .67& .72\\
 77& .52& .55& .57& .26& .12& .11\\
 75&&&&\ 0.&-.27&-.28\\
 55& .45& .41& .34&-.31&-.22&-.20\\
 53&&&&.46& .50& .48\\
 33& .37& .35& .28& .16& .03& .01\\
 31&&&& 0 &-.21&-.18\\
 11& .26& .17& .13&-.26&-.22&-.20\\
\tableline
$E_{\Gamma(4)}$&&-2.65&& &-4.78&\\
\tableline
$ E_{\Gamma(3+4)}\over 2$&&&-2.76& &&-5.06\\
\tableline
\end{tabular}
\end{table}
\begin{table}
\caption{Energies and eigenstates of the dominant terms
($\gamma=21$ added for
illustrative proposes). See text.}
\begin{tabular}{ccccccccccc}
\tableline
&$\gamma$&$e_3^\gamma$&$e_4^\gamma$&$e_{3+4}^\gamma$&M&
$<u_3| M>$
&$<u_4| M>$&$<u_3| u'_1>$&$<u_4| u'_2>$&$\alpha^2$\\
\tableline
&11&1.77&2.01&3.90&$\sigma\tau$&.992&.994&.999&1.000&.94\\
&20&-1.97&-2.14&-3.88&$r^2Y_2$&.996&.997&1.000&1.000&.95\\
&10&-1.02&-0.97&-1.96&$\sigma$&.880&.863&~.997&~.994&1.04\\
&21&-0.75&-0.85&-1.60&$r^2Y_2\tau$&.991&.998&~.999&~.997&.94\\
&&&&&&&&&&\\
&$\Gamma$&$E_3^\Gamma$&$E_4^\Gamma$&$E_{3+4}^\Gamma$&P&$<U_3| P>$
&$<U_4| P>$&$<U_3| U'_1>$&$<U_4| U'_2>$&$\alpha^2$\\
& 01&-2.95&-2.65&-5.51&$P_{01}$&.992&.998&1.000&~.994&1.048\\
&10&-4.59&-4.78&-10.12&$P_{10}$&.928&.910&~.998&~.997&~.991\\
\tableline
\end{tabular}
\end{table}

Interesting as these results might be, the truly remarkable ones come when
we diagonalize in {\it two} major shells. Let us go back to table~I, and
note that the eigenstates can always be written as
\begin{equation}
U_{3+4}=(\alpha U'_3+\beta U'_4)/\sqrt2\qquad \alpha^2+\beta^2=2,
\label{ab}
\end{equation}
where $U'_p$ can - in principle - be any unit vector, but in fact it is
almost identical to $U_p$. This is always the case as table~II shows:
 $<U_i| U'_i>$ and  $<u_i| u'_i>$ are strikingly close to 1 with
 no exceptions while $\alpha^2$ is quite close to 1.

 Therefore, for any {\it normalized} pairing or multipole operator
 $\bar O$ we have that

\medskip
{\it If $\bar O_p$
 and $\bar O_{p+1}$ are eigenstates for shell p and p+1 separately, then
$(\bar O_p+\bar O_{p+1} )$ is very much the eigenstate for the space
of the two shells. The eigenvalues are very close  in the three cases.}
\medskip

Note that we have chosen a normalization of 2 for the two shell eigenstates
so as to halve its eigenstate.

Before we examine the consequences of this result, we mention a few facts
about the other contributions.

 In table II we have added  the $\gamma=21$
case as a reminder that isovector multipoles are always present. Their
strength is between 30 and 40\% of that of the isoscalar terms (precise
numbers are given in table~III in section~II)  and they have
identical structure.

The dominant negative parity contributions are

$\gamma=10$ at 4.59$\,$MeV, $<rY^1| u^{10}> =0.994$

$\gamma=30$ at 2.69$\,$MeV, $<r^3 Y^3| u^{30}> =0.986$

The first is a center of mass operator. Its presence simply reflects the
translation invariance of the interaction. Its
$\gamma= 11$ counterpart, associated to the giant dipole resonance (GDR),
 comes at $1.81\,$MeV.
The other strong term is responsible for octupole collectivity.
(In deciding whether a given multipole is attractive or repulsive it should
be remember that $(M^\gamma M^\gamma)^0=(-)^\gamma[\gamma]^{-1/2}
(M^\gamma\cdot M^\gamma$)).

\smallskip

{\bf{d) Universal scaling}}.
We have now the necessary elements to construct a schematic but accurate
collective Hamiltonian. From \cite{ACZ91} we know that

\begin{eqnarray}
W_{xy}^\Gamma(\omega)\cong {\omega\over\omega_0}
 W_{xy}^\Gamma(\omega_0)\label{6}
\end{eqnarray}
and therefore the eigenvalues in table~II must scale in the same way.
 Setting  $\alpha^2=1$ in eq.(\ref{ab}) for simplicity,
 the {\em normalized} pairing and quadrupole forces become
   \begin{eqnarray}
H_{\bar P}=-{\hbar\omega\over\hbar\omega_0}| E^{01}|
(\overline P\,^+_p+\overline P\,^+_{p+1})
\cdot(\overline P_p+\overline P_{p+1})\label{7a}\\
H_{\bar q}=-{\hbar\omega\over\hbar\omega_0}| e^{20}|
(\bar q_p+\bar q_{p+1})
\cdot(\bar q_p+\bar q_{p+1}).\label{7b}
\end{eqnarray}
which we take as representative of the ``collective'' Hamiltonian because of
their known coherence. For the  other strong terms the expressions are
strictly similar, and all arguments concerning pairing and quadrupole
 expressions apply to them.
Since
$e_1^\gamma\cong e_2^\gamma\cong e_{1+2}^\gamma/2$ (same for $E^\gamma$),
 the coupling constants could be taken to be
independent of the space chosen; which may be any of the shells or the
two together.

\medskip

The term {\em normalized} applies to the one shell operators. For
two or more shells it is more convenient {\em not} to normalize their sum,
{\em since it is in this form that the couplings are constant}.

\medskip
 Equations (\ref{7a}) and (\ref{7b}) call for a generalization to an arbitrary
number of shells and it does not take much imagination to discover what it
should look like. However it is interesting to understand the origin of this
very welcome simplicity.
Let us  start from the obvious it is always possible to write the lowest state
 associated to multipole $\gamma$ as
\mbox{$\sum {\cal O}^\gamma_p/{\cal N}^\gamma_p$.}

To say something rigorous about the radial form of ${\cal O}^\gamma_p$
does not seem trivial, but granted that it is  close to the multipole
 operator in shell $p$, then ${\cal N}^\gamma_p$ is fixed by a scaling
argument.

In leading order the expectation value of a Hamiltonian must go as the number
of particles in the system. Therefore, in ${\cal H}$, the leading monopole
terms
must go as $O(A)$, and ${\cal H}_M$ acting on one shell must go at most
as $O(D=O(A^{2/3})$, where $D$ is the degeneracy of the Fermi shell. Since
there are $p$ possible contributors, each individual multipole term
must go as $O(A^{1/3})$.

\medskip

 {\em It is precisely what the normalized operators ensure, given
the universal scaling provided by the $\hbar\omega$ factor}.

\medskip

In our particular cases: $n$ particles in shell $p$ can produce an energy
proportional to $nD$ for the pairing, and $n^2D$ for the quadrupole
forces, which the normalizations reduce to $n$ and $n^2/D$.
Since $n=O(D)=O(A^{2/3})$, when multiplied by $\hbar\omega =O(A^{-1/3})$
both contributions become $O(A^{1/3})$.
If the n particles are promoted to some higher
shell with $p=p+M$, there is no pairing gain, a slight quadrupole loss
and a monopole loss $O(M\omega)$
that remains $O(A^{1/3})$ as long as $M$ is not too large.

For the conventional pairing and quadrupole forces the energies in one shell
would be $nD/A$ and $n^2D/A^{5/3}$ respectively, as their scalings are chosen
to yield the correct $O(A^{1/3})$ magnitude. But now when the particles
are promoted $D\rightarrow (p+M)^2$, and not only there is a gain: it becomes
larger than the monopole loss that is only linear in $M$ and the system
collapses. If the forces are restricted to act in finite spaces, to obtain
 sensible results the coupling constants must be reduced as the space
 is increased. (The problem will be discussed in Appendix B).

\medskip

The normalization is defined only to leading order $O(p^k)$, and we can
say nothing about
the $O(p^{k-1})$ terms, responsible for $\alpha^2\neq 1$, and the slight
differences we have neglected in the $e$ and $E$ couplings, but this is
the good reason to neglect them.

 What the argument does not explain is why the pairing and multipole
 operators in one shell resemble their ideal counterparts as closely as
 we have found so far. There is a hint, however, in that they must be the
 ones capable of producing coherence. More on this on the next subsection, and
  in section III, it will be shown that the $q_p'$
that couples shell $p$ to $2\hbar\omega$ jumps, remains indeed close to
$q_p$.

\medskip

We have found therefore, that we can recover the geometrical
simplicity of the pairing plus quadrupole model
without its fundamental flaw: the space dependence of the coupling constants.
The model has an enormous historical interest, and it is very instructive to
show how far we can go in justifying it (see Appendix B).

\smallskip

{\bf e) The monopole hint.} If ${\cal H}_M$, is as good as we have argued
why not trust the information it can provide about ${\cal H}_m$? It is quite
possible that, rather than wrong, it is only insufficient and its study is
most interesting.

 As befits the leading term in a multipole expansion, the monopole one is
the strongest: in fig.~2 it would come at $-10$ MeV. Unsurprisingly it has the
form $\sum n_p/\sqrt{D_p}$, which is what we expect of normalized operators,
but it should be mentioned that there are several monopole candidates one
can think of: $1$, which is the $\hat n$ operator in second quantization; but
also $r^2$ for instance. When normalized they give the same result.

The remarkable thing about this form is that it provides the answer to the
problem raised at the end of the preceeding paper. Which is that
$\hat n\hat n$, suffers from the defect of the conventional separable
forces in that it must be associated with a coupling constant that is space
dependent. There is no collapse now because $n$ is a conserved quantity, and
an $A^{-1}$ scaling ensures the asymptotically correct behaviour and the
coupling tends to a constant. Contrary to the pairing and quadrupole cases
there can be no energetic gain in promoting particles to higher orbits but
there is no loss either, and this is a subtle form of collapse because a
good  Hamiltonian
{\em must} ensure the existence of a Fermi level, i.e., it must force the
particles to occupy the lowest orbits.
The normalized monopole operator does it by producing a discontinuity
at each shell closure. Therefore, it is not only responsible for the bulk
energy of nuclear matter, but it also takes care of the major shell effects.

This operator suggests the starting point in the construction of
${\cal H}_m$. For a preliminary attempt see
 \cite{DZ95}, where a mass formula
- of rather high precision by present standards - is derived.

\medskip

\section{ Choice and complete extraction of $ H_C$}

{\bf The choice of $H_C$.} The results so far invite a separation:
\[H=H_m+H_C+H_R\]
where $H_C$ is the collective or coherent part, while $H_R$ is the rest.
To define them with some precision , we shall rely on
 the result of Mon and French \cite{MF75} that $H_R$ could be viewed
as  random , as soon as its $E$ distribution becomes symmetric.

The distributions will be characterized by their moments:
\[m_k={2\over D(D-1)}\sum E_{\Gamma}^k, \;\; \gamma_3=m_3/m_2^{3/2}\]
and the vanishing of the skewness $\gamma_3$  will be seen
 to be sufficient to ensure symmetry. We use $m_2=\sigma^2$.

Let us then  define some cut off $\varepsilon$,
eliminate from $H$ in the $e$ representation those states with
$| e_\gamma| > \varepsilon$, and decrease $\varepsilon$ until
the $E$ distribution becomes symmetric.
The result of the operation is shown in figs.~3,4 for $\varepsilon=
$2 and 1.3 MeV respectively. Labelling the original distribution in fig.~1 as
 $(\varepsilon=\infty)$, we find the following moments ($\sigma^2$ in MeV)

\begin{table}
\caption{The $sdg+pf$ states in the $f$-representation
 $(| e_\gamma|>1.3)$
and their $pf$ and $sdg$ counterparts. ${\cal S}^\pi,\,{\cal A}^\pi=$
symmetry type and
parity. $\Uparrow $ signals the states singled out in fig.2, and their one
shell counterparts. $\downarrow$ is for states with $|e_\gamma|<1.3$MeV that
are likely to have a clear multipole character. Parenthesis indicate that
the assignment is unchecked but given as plausible.}

\begin{tabular}{|c|cc|cc|c|cc|}
\tableline
$\lambda\tau$&sym& type&$e_{pfsdg}$&&$e_{pf}$&$e_{sdg}$&\\
\tableline
 00&${\cal A}^-$&$ \ldots  $&1.88&&&\\
 10&${\cal S}^-$&$  r_1Y_1 $&4.59& ${\Uparrow}$   &&\\
   &${\cal S}^+$&$  \sigma $&-1.96& ${\Uparrow}$  &-1.02&-0.97&  ${\Uparrow}$
\\
   &${\cal S}^-$&$  \ldots $&-1.53&&&&\\
   &${\cal S}^+$&$  \ell   $& 1.44&&0.66&0.80&\\
   &${\cal S}^-$&$  \ldots $&1.41&&&&\\
 11&${\cal S}^+$&$\sigma\tau$&3.90& ${\Uparrow}$   &1.77&-2.01& ${\Uparrow}$ \\
   &${\cal A}^-$&$  \ldots $&-1.83&&&&\\
   &${\cal S}^-$&$  r_1Y_1\tau$&1.81&&&\\
 20&${\cal S}^+$&$  r^2Y_2$&-3.88& ${\Uparrow}$   &-1.97&-2.14& ${\Uparrow}$ \\
   &${\cal S}^+$&$  \ldots$& 1.31&&0.64&0.75&\\
 21&${\cal S}^+$&$  r^2Y_2\tau$&-1.60&&-0.75&-0.85&\\
   &${\cal S}^-$&$  M2??$&-1.55&&&&\\
   &${\cal A}^+$&$  \ldots$& 1.46&&0.64&0.76&\\
 30&${\cal S}^-$&$  r^3Y_3$&2.69&  ${\Uparrow}$ &&&\\
 31&${\cal S}^-$&$  r^3Y_3\tau$& 1.14&$\downarrow$&&\\
 40&${\cal S}^+$&$ (r^4Y_4)$&-2.11& ${\Uparrow}$ &-1.12&-1.24&${\Uparrow}$\\
 41&${\cal S}^+$&$ (r^4Y_4\tau)$&-0.91& $\downarrow$ &&\\
 50&${\cal S}^-$&$ (r^5Y_5)$& 1.75&&&&\\
 51&${\cal S}^-$&$ (r^5Y_5\tau)$& 0.78&$\downarrow$&&&\\
 60&${\cal S}^+$&$ (r^6Y_6)$&-1.26&$\downarrow$\,&-0.73&-0.82&\\
\tableline
\end{tabular}
\end{table}

\begin{itemize}

\item
$\varepsilon=\infty \qquad \sigma^2=.99,\;\;\gamma_3=-2.22$.

 Since the lowest state
$\Gamma=1^+0$ is at $-10.12\,$MeV its contribution to $\gamma_3$ is by far
the largest but still only $-0.61$.
It means that many states in the tail must contribute to $m_3$.
\item
$\varepsilon=2.0 \qquad \sigma^2=0.60,\;\;\gamma_3=-0.79$.

Five peaks have been excluded
$\gamma=1^-0,\,1^+1,\,2^+0,\,3^-0$ and $4^+0$
and now the
$\Gamma=1^+0$ state at $-7.79\,$MeV accounts for most of $\gamma_3$
with a contribution of $-0.60$, which when added to that of the next
 two states
$\Gamma=2^+0$ ($-4.29\,$MeV) and $0^+1$ ($-3.82\,$MeV),
at $\gamma_3=-0.86$, exceeds the full value.
 As anticipated, the very large $e$ states are
responsible for most of the tail of the original distribution.
Although distorsions are still apparent, they may be interpreted as
fluctuations and we have a good model for $H_C \equiv$ the five $\gamma$
states and the  $\Gamma=1^+0,\,0^+1$ ones.

Since it is not only the size, but also the ability to generate coherence
that must characterize the main terms we have left out $\Gamma=2^+0$
which - as we shall see in section IV - must be counted as a nuisance,
rather than a bona fide candidate to $H_C$.

It should be noted that the structure of the large
$\Gamma$ states is little changed in going from $\varepsilon=\infty$ to
$\varepsilon=2$.

\item
$\varepsilon=1.3 \qquad \sigma^2=0.41,\;\gamma_3=-0.002$

 This cut-off was chosen as a sensible
definition of the bulk of the $e$ distribution. The histogramm in fig~4
now becomes structureless.
The $\gamma$ states with $|e_\gamma|> 1.3$ are listed in table~3.
 Some useful information is given, as their symmetry type (see eqs. (1.30-32)
 in [1]), and their multipole nature whenever possible.
\end{itemize}

\bigskip

The cut off is reasonable in that the majority of excluded states
belong naturally to what is expected from a multipole decomposition.
 However, several peaks have an ill defined status, while others
we would include in $H_C$ have missed the cut off (e.g. $3^-1$ at $e=1.14$).
Clearly we are in the boundary where the small terms in $H_C$
overlap the large ones in $H_R$.

Two attitudes compete in choosing $H_C$: Include either {\it as much
as necessary} or {\it as little as possible}. Which one is given precedence
depends on the computational strategy adopted.

\medskip

{\it As much as necessary} applies to the recently developped
shell model Monte Carlo approach \cite{ALHA}. The Hamiltonian must
be written in some $e$ representation, and the authors have chosen an
ingenious one in which they introduce the Pauli violating terms necessary
to cancel all $\gamma=\lambda1$ terms. To understand how it can be done
refer to eq.(1.47b) in the precceding paper[1], and imagine that in
$V^{JT}_{rstu}$ the $rs$ and $tu$ states are allowed to be symmetric.
Since the purpose of the exercise is to reduce the number of terms, it is
probably a good idea to proceed as we have done and choose a proper $H_C$.
The problem is that pairing has to be included in the $e$ representation
which may be costly, but an example given at the end of the section suggests
that the cost is probably not too high.
\medskip

{\it As little as possible}, can be reconciled with as much as necessary
by treating first the truly important terms and then doing some form
 of perturbation theory to account for the rest.  Here
 pairing plus quadrupole or even only one of them may be sufficient.

\smallskip

One example may be mentioned. In a forthcoming study \cite{ZRPC95},
it is shown that states of 4 neutrons and 4 protons moving in the
same or contiguous major shells, generate rotational spectra exhibiting
systematic backbending. With
a proper $\bar q\cdot \bar q$ the wavefunctions have overlaps of better than
0.95 with those of the realistic interaction $H$, but the backbending has gone.
However when the expectation value of the full $H$ wis calculated using the
$\bar q\cdot \bar q$ eigenstates, the spectrum reproduces perfectly the exact
one: lowest order perturbation theory may work very well with a reasonably
good starting vector.

\smallskip

This example is particularly relevant because the ``proper'' quadrupole that
reproduces best the exact results is some 30\% larger than $e$ eigenvalue
in table~3. It is probable that part of the effect is due to the isovector
quadrupole that was not included in the calculation. But there is another
cause as we see next.

\medskip

{\bf Complete Extraction}

\medskip

 $H_C$ and $H_R$ in the  normal representation, can be thought as
 vectors, each matrix element $W_{rstuC}^\Gamma$ and $W_{rstuR}^\Gamma$,
 being associated $[\Gamma]$ to unit vectors. Then  the norm
  is simply $m_2\equiv \sigma^2$ (assume in all that follows that the
 matrix elements have been all divided by $D(D+1)/2$)
\begin{eqnarray}
\sigma^2=<(H_C+H_R)^2>=\sum (W_{rstu}^\Gamma)^2[\Gamma]
\nonumber\\
=\sum\left[ (W_{rstuR}^\Gamma)^2+(W_{rstuC}^\Gamma)^2
+2W_{rstuR}^\Gamma W_{rstuC}^\Gamma\right][\Gamma]\nonumber\\
\equiv H_C^2+H_R^2+2H_C\cdot H_R
\label{V.14}
\end{eqnarray}
It is elementary to show that if $H_1$ and $H_2$ are the Hamiltonians
associated to eigenvectors $E_1$ and $E_2$, then
 \mbox{$<H_1H_2>=H_1\cdot H_2=0$.}

The crux of the matter is that $H_C$ and $H_R$ are by construction orthogonal
in the multipole representation, but they need not be in the normal one. We
may even invent vectors orthogonal in the former that become parallel
in the latter. This, because of the non linear independence problem
 encountered several times in [1]: a multipole decomposition can teach us
 {\it what} to separate, but not {\it how} to separate.

 What has been done in the preceeding section illustrates quite well
 the first part of the proposition, and incidentally the need to use
 the $f$ matrix elements in eqs.(1.2), to ensure symmetric factorizations,
 since any other choice would have led to opaque results. Now that we know
 what to separate, we should do it well.

 The simplest way is to write

 \[H_R+H_C=\left (H_R-{H_C\cdot H_R \over H_C\cdot H_C}H_C\right )
            +H_C\left (1+{H_C\cdot H_R \over H_C\cdot H_C}\right ),\]
 which makes the two terms orthogonal. By squaring we find
 \[\left[ H_R^2-{(H_C\cdot H_R)^2 \over(H_C\cdot H_C) }    \right]+
   \left[ H_C^2+2H_C\cdot H_R+ {(H_C\cdot H_R)^2 \over(H_C\cdot H_C)}\right]\]
an since ${(H_C\cdot H_R)^2 \over(H_C\cdot H_C) }$ can be safely neglected,
the orthogonalization amounts to leave $H_R$ almost untouched, and to boost
$H_C$ by the cross term. Now we call upon eq. (\ref{III.14}), which says
that the total norms in the two representations are proportional and
 examine the relative contributions of $H_C$ and $H_R$ in both cases.

For $\varepsilon=2$
, from table~3 we can find that the five largest peaks contribute .29 ,
i.e. some 30\% to the total norm which we know to be .99, and therefore
$H_R$ accounts for the remaining 70\%.
But we also know that - when calculated in the normal representation -
$H_R^2$ is not .70, but .60, and therefore the orthogonalization boosts the
share of $H_C$ to 40\%, which implies a $\sqrt{4/3} \equiv 15\%$
increase in the coupling constants.

This should be sufficient to have an idea of the effect, but for a
quantitatively reliable extraction, each term must be treated separately.
Let us see how.

\medskip

Assume we have selected the candidates to $H_C$, transformed to the normal
represetation the multipole ones, and made them monopole free. Call them
$H_\kappa$. We want to find the linear combination
 \mbox{$H_C=\sum C_\kappa H_\kappa$},
that maximizes the overlap between $H_C$ and $H$.
It amounts  to solve the standard
problem $(\alpha\equiv rstu)$
\begin{eqnarray}
\sum_\alpha (W_\alpha^\Gamma-\sum_\kappa C_\kappa W_{\alpha\kappa}^\Gamma)^2
[\Gamma] =(H-\sum_\kappa C_\kappa H_\kappa)^2&\nonumber\\
=H\cdot H-2\sum_\kappa C_\kappa H_\kappa\cdot H+
2\sum_{\kappa,\kappa'}
C_\kappa C_{\kappa'}H_\kappa\cdot H_{\kappa'}=\min\label{V.15}
\end{eqnarray}
and therefore $C_\kappa$ is determined by the linear system
\begin{equation}
H_\kappa\cdot H=\sum_{\kappa'} H_\kappa\cdot H_{\kappa'} C_{\kappa'}
\label{V.16}
\end{equation}

The $H_\kappa$ vectors can be made orthogonal by diagonalizing the norm
matrix $H_\kappa\cdot H_{\kappa'}$ through transformation $T_{\mu\nu}$
\begin{equation}
H_\mu=\sum T_{\mu\kappa} H_\kappa\,,\quad
H_\kappa=\sum T_{\kappa\mu} H_\mu\,;\quad
H_\mu\cdot H_{\mu'}=\delta_{\mu\mu'} H_\mu^2\,.
\label{V.17}
\end{equation}
Therefore, defining $H_C$ in terms of the $H_\mu$, we have
\begin{equation}
H_C=\sum C_\mu H_\mu\quad\stackrel{(\ref{V.16})}{\Longrightarrow}
\quad  C_\mu=H_\mu\cdot H/H_\mu^2\,.
\label{V.18}
\end{equation}

As we see the analysis in the first section has fufilled its part
in pointing to the correct forms to be extracted, but the associated coupling
constants may need some corrections. If our estimates are correct,
they should not exceed 20\%.

It is worth emphasizing that once the fully extracted $H_C$ is transformed
back to the multipole representation it may {\it look} quite different from
the original, which is not monopole free, and contains Pauli violating terms.
However, it is very much the original.

A potentially useful variant of this possibility was found in the $sd$ shell:
When the strongest
contribution for each $\gamma$ is included and extraction is carried
maximizing the overlap, the addition of the pairing term makes no
 difference.

\section{ Core polarization}

To account for the effect of states outside of one or two major shells
in which we are prepared to work, we use the quasiconfiguration method
\cite{PZ81a,PZ81b}. The idea is to separate the full space into a model space
containing states $|i>$ and an external one made of states  $|j>$. The $|i>$
and $|j>$ states are now ``dressed''  through a transformation that
respects strict orthogonality
\[|\bar \imath>=|i> +\sum_jA_{ij} | j>  \quad
   |\bar \jmath>=|\bar j> -\sum_iA_{ij} |i> \quad
   < \bar \imath|\bar \jmath>=0 .\]
The amplitudes $A_ {ij}$ are then defined by
  \[ < \bar \imath|{\cal H}|\bar \jmath>=0 .\]
Perturbation theory is the natural way to determine $A_ {ij}$, since
whatever must be treated exactly is in principle contained in the model space,
 which we take here to be a full shell $p$. What is nice about the
 quasiconfiguration method, is that it makes it very clear how the dressing
 of the states is transformed into the dressing of the operators.

We refer for details to \cite{PZ81a,PZ81b}, from which we borrow the
expression
 for the second order correction, and apply in some detail to the pairing,
 and quadrupole
 Hamiltonians.

\medskip

 {\bf Pairing.} We deal first with the effect of shell $p+1$ only. $E_{01}$
is a one shell energy from table~2, scaled by $\hbar \omega/\hbar \omega_0$,
and $\hbar \omega_0=9$MeV.

\begin{mathletters}
\label{P}
\begin{equation}
 <\bar \imath | H_{\bar P}| \bar \imath'>=
 < i|{E_{01}(\omega)}{P_p^+\cdot P_p\over \Omega_p}| i'>\\
\end{equation}
\begin{equation}
-{E_{01}^2  (\omega)\over 2\hbar\omega}\sum_J
{< i| P_p^+\cdot P_{p+1}| J>
< J| P_{p+1}^+\cdot P_p| i'>\over
\sqrt{\Omega_p\Omega_{p+1}}
\sqrt{\Omega_p\Omega_{p+1}}}=\\
\end{equation}
\begin{equation}
=< i| {E_{01}(\omega)} {P_p^+\cdot P_p\over\Omega_p}
-{E_{01}^2(\omega)\over 2\hbar\omega}{1\over 3}
{P_p^+\cdot P_p\over\Omega_p}
{P_{p+1} \cdot P_{p+1}^+\over\Omega_{p+1}}| i'>=\\
\end{equation}
\begin{equation}
< i| {E_{01}(\omega)\over 2}\left(1-
{E_{01}(\omega)\over 4\hbar\omega_0}\right)
{P_p^+\cdot P_p\over\Omega_p}| i'>
\end{equation}
\end{mathletters}
Step by step we find

a) the unperturbed energy, to which we add

b) the second order perturbation. The intermediate states
are assumed to be all at energy $2\hbar\omega$. Then

c) invoking closure and recoupling the operators (hence the factor 1/3),

d) the final result follows by contracting out the the operators in shell
$p+1$ (which gives back a factor 3).

Repeating the operation to account for states in shell
$p-1$ leds to exactly the same correction and we conclude that the
renormalized operator is the original one affected by a modified $E_{01}$
\begin{equation}
{E_{01}} \quad\longrightarrow\quad {E_{01}}\left(1+
{|E_{01}|(\omega)\over\hbar\omega}\right)\equiv E_{01}\longrightarrow
E_{01}(1+0.32),\label{Pres}
\end{equation}

\medskip

Now consider what would happen if we were to generalize to an indefinite
number of shells, and compare with the result for ordinary pairing:
\begin{eqnarray}
{E_{01}(\omega)} \longrightarrow  {E_{01}(\omega)}\left[ 1+
{|E_{01}|(\omega)\over 2\hbar\omega}\left(1+{1\over 2}+{1\over 3}+\cdots
\right)\right]\label{PP}\\
G \longrightarrow  G\left[ 1+
{|G|\over 2\hbar\omega}\left(\Omega_{p+1}+{\Omega_{p+2}\over 2}+
{\Omega_{p+3}\over 3}+\cdots
\right)\right].
\label{PPP}
\end{eqnarray}
The normalization of the operators $P_p$ transforms
a bad divergence into a logarithmic one.
Since high lying pairing excitations violate badly translation
invariance the corresponding matrix elements should be quenched,
and this is probably sufficient to eliminate the divergence.
Plausible as this argument may be, the fact remains that the
renormalization of the pairing force is a delicate problem
that has yet to be solved. Some more on this at the end of the section.

\medskip

{\bf Quadrupole.}
The problem of multipole renormalizations is very different from the one we
have analyzed for pairing because the physical processes are different.
Where we had two particles jumping to higher shells, or being promoted
from lower ones, now it is a particle-hole excitation that is produced by
the valence particles, and it is in the multipole representation that the
problem is naturally treated, and the space involved is always finite in
second order for the operators of interest.

In particular, the quadrupole renormalization are due to $2\hbar\omega$ jumps
mediated by a term of the form
\mbox{$k(\bar q\,'_p\cdot\bar q\,'_{2\hbar\omega}
+\bar q\,'_{2\hbar\omega}\cdot\bar q_p)$},
which must
be extracted from ${\cal H}_M$. The primes indicate that the operators
may not be genuine quadrupoles. The extraction method is a variant
of the diagonalizations we have used so far. It is explained in  Appendix A
under the item {\it asymmetric factorization}. The space involves all shells
from $p=0$ to 4, and the matrix elements
interested in are eliminated (those of the form
\mbox{$\bar q\,'_{2\hbar\omega}\cdot \bar q\,'_{2\hbar\omega}$})
 Note that there is no cheating here: we could have done the
same thing in the two shell case by keeping only the cross shell elements
if we had wanted only  the $\bar q_p\cdot \bar q_{p+1}$
component. To compare with empirical
reults in the $sd,\;p=2$ shell we work at $\hbar\omega=11$. The standard one
shell calculation yields $e^{20}=-2.40\,$MeV (note that eq.(\ref{qnum}) gives
 $-2.37\,$MeV). For the cross terms the large matrix produced
\begin{eqnarray}
k=-2.80\,{\rm MeV},\quad<\bar q\,'_2|\bar q_2>=0.97,\label{imp}\\
<\bar q\,'_{2\hbar\omega}| q_{02}+q_{13}+q_{24}>/{\cal N}_{2\hbar
\omega}=0.83\label{83}
\end{eqnarray}
where $q_{pp'}$ are the $2\hbar\omega$ quadrupole operators that can
couple to $q_2$. Their sum is normalized to ${\cal N}_{2\hbar\omega}$.
By normalizing each $q_{pp'}$ operator separately the overlap at 0.87 is
slightly better, but it is the number in (\ref{83}) we shall need.

{}From (\ref{imp}) $\bar q\,'_2\approx \bar q_2$, and this is an important
message: it does not matter much to what $\bar q_2$ is coupled - even
$4\hbar\omega$  jumps if ${\cal H}$ demands them (not very likely) - but
it must be $\bar q_2$ to ensure that  ``quadrupole renormalizes quadrupole''
and that effective charges are state independent as shown below.

Proceeding as in eqs.(\ref{P}) ,
 we have ($hc$ stands for
Hermitean conjugate of the first term)

\begin{mathletters}
\label{qua}
\begin{equation}
<\bar \imath| H_{\bar q}|\bar \imath'>
={e_{20}(\omega)}
< i| \bar q_p\cdot\bar q_p| i'>-\\
\end{equation}
\begin{equation}
{k^2(\omega)\over 2\hbar\omega}
< i|( q' _{2\hbar\omega}\cdot \bar q_p+ hc)|J><J|
( q'_{2\hbar\omega}\cdot \bar q_p+ hc)| i'> =\\
\end{equation}
\begin{equation}
\cong<i| {e_{20}(\omega)}
 \bar q_p\cdot\bar q_p -
{2k^2(\omega)\over \hbar\omega} {1\over 5}
  q'_{2\hbar\omega}\cdot
 q'_{2\hbar\omega}\bar q_p\cdot \bar q_p| i'> =\\
 \end{equation}
\begin{equation}
 =\left({e_{20}(\omega)}-{k^2(\omega)\over \hbar\omega}\right)
< i| \bar q_p\cdot\bar q_p| i'>
\end{equation}
\end{mathletters}
 In the last step  only half of the operators
act: those creating first an exitation and then destroying it. Hence the factor
5/2 instead of 5.
  We have indulged in the fallacious
approximation of treating the three terms in
$q'_{2\hbar\omega}\equiv q'_{p-2p}+q'_{p-1p+1}+q'_{pp+2}$ as
commuting with $\bar q_p$, which is true only for the middle one. A
correct calculation would yield a rank $2+3$ force for the offending terms.
Still the result is correct for the two body contribution as can be checked in
\cite{PZ81a} in which a very similar case is fully worked out. The neglect of
three body contributions is common, but bad, practice.
\medskip

The modification of the transition operator $q_p$ (not $\bar q_p$) is
calculated along similar lines,

\begin{mathletters}
\label{eff}
\begin{equation}
<\bar i| q|\bar i'>=
< i|  q_p| i'>-
{2k(\omega)\over 2\hbar\omega}
< i| q_{2\hbar\omega}
( q'_{2\hbar\omega}\cdot \bar q_p+ hc) | i'>
\end{equation}
\begin{equation}
=< i| q_p| i'>-{2k(\omega)\over \hbar\omega}
 {1\over 5}{{\cal N}_{2\hbar\omega}\over{\cal N}_p}
< i| q_p \bar q_{2\hbar\omega}\cdot q'_{2\hbar\omega} | i'>
\end{equation}
\begin{equation}
=\left(1-0.83{k(\omega)\over  \hbar\omega}
{{\cal N}_{2\hbar\omega}\over{\cal N}_p}\right)
< i| q_p| i'>
\end{equation}
\end{mathletters}
what has been done is:

a) use the second order expression.

b) upon recoupling collect 4 equal terms and the factor 1/5. Then,
  to contract the normalized ${2\hbar\omega}$ operators
 (whose ovelap is 0.83), interchange normalizations
for $q_p$ and  $q_{2\hbar\omega}$, and write

c) the final result.

Inserting the necessary numbers
(${\cal N}_{2\hbar\omega}/{\cal N}_p=1.97$ for $p=2$), the transition
operator is boosted by a factor (1+0.42), which is too small to agree
with the empirical value close to 2. The problem is that
 second order perturbation may work very well for the energy
but poorly for the transition rates \cite{PZ81a}.
We shall not go into the reasons, and simply borrow a nice argument
due to Mottelson (\cite{MO60}). The idea is
 that the second order
perturbation should not be understood as affecting only the valence
particles but the whole system. Then, calling $Q^{20}$ the total quadrupole
operator ,the estimate $ Q^{20}=(1+0.42)q^{20}$ becomes
\[Q^{20}=q^{20}+0.42 Q^{20}\longrightarrow Q^{20}=(1-0.42)^{-1}q^{20},\]
a result equivalent to an RPA resummation.

The comparison with empirical data is rewarding:( $\eta^\gamma$ are
the effective charges)
\begin{eqnarray}
{e^{20}}\longrightarrow e^{20}(1+0.3)=-3.12\quad vs\;-3.18\cite{WI84}\\
\eta^{20}= (1-0.42)^{-1}=1.76 \;vs\; 1.78(3)\;\cite{BW88}\\
\eta^{21}= (1+0.19)^{-1}=0.8\;\;vs\;\; 0.8(1)\;\cite{BW88}
\end{eqnarray}
(The isovector effective charge can be calculated knowing k=1.28)

There is perhaps an element of luck here, but the presence of the overlap
and the norms in the last equation (\ref{eff}) indicates that the diagonal
representation is probably incorporating some important effects
neglected in previous calculations of the effective charges \cite{SM91}.

\medskip

The comparison of the effective pairing numbers in (\ref{Pres}), with the
empirical values would also be satisfactory, but fortuitous, because of
the logarithmic divergence of second order theory.

This problem does not seem to have attracted much attention, but there have
been numerous controversies in the literature on the influence of core
polarization in the strict sense - i.e. what we call multipole processes -
on pairing renormalization (see \cite{MOK} for a recent review).

According to our simple ideas ``pairing renormalizes pairing'' and
``multipole renormalizes multipole'' and the controversies would seem
without object, but there is a catch.

A pairing force
can always be written in the multipole representation, the highest multipoles
entering with greater weights. In the $sd$ shell - the region that generated
the controversies -, quadrupole and hexadecapole are already high
multipolarities,  well represented
in the pairing decomposition, and their strong polarizabilities will have
a non negligible effect. This is simply a manifestation of the lack of
linear independence we have often encountered.

In heavier nuclei the important contributions to $H_C$ will become
increasingly linear independent, and they will increasingly renormalize
themselves, and the question will remain: what is the precise nature of
pairing renormalization?

Whatever the answer to the question, it is clear that the use of diagonal
representations simplifies considerably the calculation of effective operators
 by making it possible to concentrate
attention on the few important terms that grow bigger, and neglect
 the small ones that have no chance to grow.

\section { Comparing interactions}.

Now we compare several realistic interactions  in the $sd$ shell, the
 only region in which a direct fit to the data, $W$ \cite{BW88}, leads
 to more accurate spectra.

Table 4 shows some off diagonal matrix elements  at the beginning
of the shell, calculated with the KLS (as used in \cite{ACZ91}),
KB \cite{KB66}, and Bonn~B \cite{JMSK92} realistic interactions, or taken
from the
$W$-fit. It is apparent that the realistic values are close to one another
(especially
KB and Bonn B) and not far from $W$, except in two cases.

 The comparison of individual matrix elements can be misleading, because
 apparently
 large discrepancies may be of little consequence, and apparently small ones
 disastrous. Therefore it is better to concentrate on the contributions
 that can
 make a difference. Accordingly table~5 collects information on the
 lowest state in the $e$ and $E$ representations. The KB numbers have been
 left out because they practically duplicate those of Bonn B.
  The differences in eigenvalues between
 KLS and Bonn B is not large but may have some significance, while
  the agreement in eigenvectors is complete.
Very much the same is true in the comparison
of $W$ with the realistic values except for $\Gamma=20$ and 30, which
are {\it not among the dominant states of the collective Hamiltonian}.
The reading of the result is simple: $W$ has discovered and dealt with some
local problem by altering radically some special matrix elements, but
{\it it has scrupulously respected the truly important contributions}.

\medskip

The local problem has several manifestations. In ${}^{22}$Na the realistic
interactions produce a close doublet $J=1$,~3, instead of the well detached
$J=3$ observed ground state\cite{BW88}, the $\gamma$ band in ${}^{24}$Mg is
too low, and more interesting for our purpose, the $JT=20$ state in ${}^{18}$F
always too low by about 1~MeV. (It should be noted that the experimental
counterpart of this level is the second $JT=20$  in ${}^{18}$F, the first
is an intruder).

 The solution proposed by the $W$ fit is such that there is no hope to
reconcile it with the realistic forces. The problem is then to reconcile
the latter with the data. We do not know how to do that, and we close
this section by looking for some clues.

\medskip

Though realistic interactions are similar, they also differ in some
respects as can be gathered from fig.~3 of ref.\cite{JMSK92}, that shows
different calculated spectra of ${}^{18}$F and ${}^{18}$O.
There are two effects.
\begin{itemize}
\item
Overall Dilation. If we refer to eq.(\ref{6}) the discrepancy can be absorbed
by simply changing slightly the oscillator frequency, which is in principle
fixed by the size of the nucleus. Because of their bad saturation properties
this choice is not always possible for the realistic potentials. As long as
saturation properties have to be treated phenomenologically, the overall
dilation must be treated as a free parameter.
\item The $JT=01$ and $10$ matrix elements. The differences due to overall
dilation can be eliminated by normalizing the interactions to the same
$\sigma^2$ in eq.(2.1). When this is done the most notable discrepancies come
from the  $JT=01$ and $10$ matrix elements, that are spectroscopically the
lightest in the sense that $[JT]$ is smallest. As we have seen in the
preceeding section they are also the ones most severely affected by
renormalization uncertainties.
\end{itemize}

Whether the local problem in the $sd$ shell can be solved by playing on
these different elements is an open question, but the analysis in terms of
the diagonal representations makes clear that there is no much room for
tampering with the dominant terms in the interactions.

\appendix
\section{}
{\bf Properties of the f matrices}

There is nothing special about the $W$ matrix that is diagonalized in the $E$
representation except that it is traceless by construction, while the $f$
matrix leading to the $e$ representation has a number of non trivial
properties.

{\it Direct properties}
\begin{itemize}
\item The $f^{\lambda\tau}$ matrix
has twice as many elements as the $W^{JT}$ matrix with $J=\lambda$:
\hfill\break
if $r\not= s$ and $t\not= u$, $W_{rstu}^\Gamma$ goes into
$\omega_{rtsu}^\gamma$ and  $\omega_{rust}^\gamma$ and the allowed values of
$\Gamma$ and  $\gamma$ are the same.

if $r= s$ or $t= u$, $W_{rstu}^\Gamma$ goes into
$\omega_{rtsu}^\gamma=\omega_{rust}^\gamma$, but
$\Gamma$ is allowed for $(-)^\Gamma=-1$ only and there is no restriction on
$\gamma$. E.g.: there are 5 possible ways of constructing $J=2$ states in the
$sd$ shell $d_{5/2}^2,\,d_{3/2}^2,\,d_{5/2}s_{1/2},\,
d_{5/2}  d_{3/2}  ,\,s_{1/2}d_{3/2}$ and there are five $JT=21$ states
and three $JT=20$ states (the first two are Pauli forbidden). For the
$\gamma$ matrices there are 8 possible combinations: the 5 above plus
$s_{1/2}  d_{5/2},\,d_{3/2}d_{5/2}$ and
$d_{3/2}  s_{1/2}$ which are counted as different.
Therefore in $W$ we have a $3\times 3$ ($\Gamma=20$) and a
$5\times 5$ ($\Gamma=21$) matrix; while in $f$ we have twice an
$8\times 8$ matrix. Each of these
$8\times 8$ matrices consists of two blocks of
$3\times 3$ and $5\times 5$ because of the following property.

\item In the $e$ representation we can write
\begin{eqnarray}
H={1\over 2}\sum_{r\leq t\atop s\leq u \,\gamma}[\gamma]^{1/2}
\left[(f^\gamma_{rtsu}+(-)^{s-u}
f^\gamma_{rtus})({\cal S}_{rt}^\gamma{\cal S}_{su}^\gamma)^0+
(f^\gamma_{rtsu}-(-)^{s-u}
f^\gamma_{rtus})({\cal A}_{rt}^\gamma{\cal A}_{su}^\gamma)^0
\right]
\end{eqnarray}
in terms of the ${\cal S}$ and ${\cal A}$ operators defined in eq.(1.31)
of [1].
 In the preceding
example it means that out of 8 possible
$\gamma=2\tau$ combinations 3 are of ${\cal A}$ type and 5 of ${\cal S}$
type.
Note that the matrices are
always symmetric but the eigenvectors are linear combinations of
 states that are either of ${\cal S}$ or ${\cal A}$ type.

\item The trace of the $f$ matrix {\it always} vanishes, as seen from
\begin{eqnarray}
\sum_{\gamma, rt} f_{rtrt}^\gamma [\gamma] =
\sum_{rt} P_{rr}P_{tt}
\sum_{\gamma[\Gamma]}(-)^{r+t-\gamma-\Gamma}
\left\{ \begin{array}{ccc}
   r&r&\Gamma\\  t&t&\gamma
   \end{array}\right\} W_{rrtt}^\Gamma=\nonumber\\
\sum_{rt}{P_{rr}P_{tt}\over [ rt]^{1/2}}W_{rrtt}^{00}=0
\end{eqnarray}
where we have used (\ref{3a}) and sum rule (A.19) from [1]. The
matrix element $W_{rrtt}^{00}$ is always zero because it violates Pauli.

\item The variance of the $f$ matrix is
$\sigma_f^2={D-1\over 8D}\sigma_W^2$, where
$\sigma_W^2$ is the variance of the $W$ matrix. The calculation is better
conducted in the $m$-scheme
\begin{eqnarray}
H(W)=\sum W_{xx'}Z_x^+Z_{x'}\nonumber\\
 x\equiv (ij)\,,\;
x=1\cdots D^{(2)}/2\,,\quad i< j \\
H(f)=\sum f_{aa'}S_aS_{a'}\nonumber\\
 a\equiv (ij)\,,\;
x=1\cdots D^2\; (ij~\rm unrestricted)
\end{eqnarray}
A typical $W_{xx'}=W_{1234}$, say, appears as
$f_{aa'}=\pm{1\over 4}W_{xx'}$ in four contributions:
$aa'=13\;24$ and 24~13, 14~23 and 23~14. Then

\begin{eqnarray}
\sigma_W^2=\left({D^{(2)}\over 2}\right)^{-1}\sum W_{xx'}^2\,;\\
\sigma_f^2=D^{-2}\sum f_{aa'}^2=\nonumber\\
D^{-2}\sum 4
\left({W_{xx'}\over 4}\right)^2={D-1\over 8D}
\sigma_W^2
\label{III.14}
\end{eqnarray}
\end{itemize}
\medskip
{\it Asymmetric factorizations}

Quadrupole renormalizations are mediated by terms of the form
\mbox{$k(\bar q\,'_p\cdot\bar q\,'_{2\hbar\omega}
+\bar q\,'_{2\hbar\omega}\cdot\bar q_p)$}. To understand how
 asymmetric factorization are possible we study the spectra of
$I\times I$ matrices $f(n)$, whose non zero elements belong to the
rectangular blocks $f_{xa}$ and $f_{ax}$, $a=1\ldots K$,
$x=K+1\ldots I,\;I-K=L$.
 \medskip

Specializing eq.(\ref{III.4}) to this situation we have
\begin{equation}
\sum_x f_{ax} u_{xk}=u_{ak} e_k\qquad
\sum_a f_{xa} u_{ak}=u_{xk} e_k\label{III.20}
\end{equation}

The eigenvector $| k>$ with eigenvalue $e_k$ can be expanded
in terms of unit column vectors $| i>$ (1 in the $i$-th position,
zero in the others):
\begin{equation}
| k>=\sum_{i=1,I} u_{ik}| i>=
\sum_{a=1,K} u_{ak}| a>+
\sum_{x=K+1,I} u_{xk}| x>
\end{equation}

By reversing simultaneously the sign of $e_k$ and $u_{xk}$
$(e_k\longrightarrow -e_k,\;u_{xk}\longrightarrow -u_{xk},\;\forall\;x)$
eq.(\ref{III.20}) remains
unchanged telling us that
\begin{equation}
|\bar k>=\sum_a u_{ak}| a>-
\sum_x u_{xk}| x>
\end{equation}

is an eigenvector with eigenvalue $-e_k$. Furthermore, from
unitarity
\begin{equation}
\sum_i u_{ik}u_{ik'}=
\sum_a u_{ak}u_{ak'}+
\sum_x u_{xk}u_{xk'}=\delta_{kk'}
\end{equation}
and taking overlaps
\begin{equation}
< k|\bar k'>= \sum_a u_{ak}u_{ak'}-
\sum u_{xk}u_{xk'}=0
\end{equation}

leading to
\begin{equation}
\sum_a u_{ak}u_{ak'}=
\sum_x u_{xk}u_{xk'}={1\over 2}\delta_{kk'}
\end{equation}

{}From these results we may construct the spectrum of $f(n)$.
Let's call $M=\min (L,K)$. We have: $M$ positive eigenvalues $e_k$
$(k=1,M)$, $M$ negative ones $e_{\bar k}=-e_k$
$(\bar k=M+k)$ and $I-2M$ null ones.
\medskip

We may gain further insight by presenting the problem as a search for an
optimum
approximant $g_a g_x$ to the rectangular matrix $f_{ax}$, defined through
\begin{equation}
\sum(f_{xa}-g_a g_x)^2=\min\,.
\label{III.26}
\end{equation}

Variation with respect to $f_x$ and $f_a$ leads to
\begin{equation}
\sum_x f_{xa}g_x=g_a \sum_x f_x^2\,,\quad
\sum_a f_{xa}g_a=g_x \sum g_a^2
\label{III.27}
\end{equation}

and given that (\ref{III.26}) is invariant
under $g_x\longrightarrow \sigma g_x\;$
$g_a\longrightarrow \sigma^{-1} g_a$ we may request
\begin{equation}
\sum_x g_x^2= \sum_a g_a^2=e
\end{equation}

and (\ref{III.27}) becomes (\ref{III.20}) by identifying $g_x=u_{xk}$,
$g_a=u_{ak}$, $e=e_k$. The factorization produced by the lowest
$(k=1)$ eigenstate at $e=-| e_1|$ is identical to the one
for the highest at $e=| e_1|$ and the best available. Exact
separability is achieved for $e_k=0$ for $k\not=1$.

\section{}
 {\bf Baranger and Kumar revisited}.

In two famous papers Baranger and Kumar attempted to derive from a realistic
interaction the pairing plus quadrupole forces adapted to a space of two
major shells \cite{BK68} and proceeded to do Hartee-Fock-Bogoliubov (HFB)
calculations in the rare earth region that showed for the first time that
it was possible to explain microscopically the onset of
deformation\cite{KUBA68}.

The success of the calculations is probably due in large part to the fact
that the model is far more realistic than its authors believed. The reason
is rather strange and we think it is worth telling.

 \medskip

Let us start by comparing the traditional pairing plus quadrupole forces
as used in \cite{BK68,KUBA68}
\begin{eqnarray}
H_P=-G_x\,(P_p^++P_{p+1}^+)\cdot (P_p+P_{p+1})
\\
H_q=-{\chi'\over 2}\,(q_p+q_{p+1})\cdot (q_p+q_{p+1})
\end{eqnarray}
with the normalized versions in eqs.(\ref{7a}) and {\ref{7b}), which
we write explicitly by borrowing numbers from table II and
remembering that $\hbar\omega_0=$9~MeV:
\begin{eqnarray}
H_{\bar P}=-0.32\hbar\omega\left({P_p^+\over\sqrt{\Omega_p}}+
{P_{p+1}^+\over\sqrt{\Omega_{p+1}}}\right)\cdot
\left({P_p\over\sqrt{\Omega_p}}+
{P_{p+1}\over\sqrt{\Omega_{p+1}}}\right)
\\
H_{\bar q}=-0.216\hbar\omega\left({q_p\over{\cal N}_p}+
{q_{p+1}\over{\cal N}_{p+1}}\right)\cdot
\left({q_p\over{\cal N}_p}+
{q_{p+1}\over{\cal N}_{p+1}}\right).\label{qnum}
\\
\end{eqnarray}

If we consider first the case of one shell, the operators are the same, and we
can relate the coupling constants by simply equating.

For the two shell case, the overlap
\begin{equation}
{1\over \sqrt2}<{q_p\over{\cal N}_p}+{q_{p+1}\over{\cal N}_{p+1}}\left|
{q_p+q_{p+1}\over\sqrt{{\cal N}_p^2+{\cal N}_{p+1}^2}}\right.>={1\over\sqrt2}\,
{{\cal N}_p+{\cal N}_{p+1}\over\sqrt{{\cal N}_p^2+{\cal N}_{p+1}^2}}=0.98\;{\rm
for~}p=3
\label{ov}
\end{equation}
seems large enough to identify $H_{\bar q}$ and $H_q$ to a good approximation.
Of course there is some risk because the norms defined in the vector space of
the multipole representation  the $S_{rt}$  operators are {\it all}
treated as unit vectors. In a midshell situation, as in ${}^{28}$Si, the
$p+1$ - upper - shell is very poorly represented in the wavefunctions.
Then the large overlap is meanigless.

With this proviso in mind we  equate the traditional and the new forms.
 Recalling that $\hbar\omega=40\,A^{-1/3}$
 \cite{BM64}, and writing the norms in terms of $A_{mp}$ as in eqs.(6,7)
 we find for one shell
\begin{equation}
{0.216\hbar\omega\over {{\cal N}_p^2}}\cong {1\over 2}\,
{216\over A^{1/3}A_{mp}^{4/3}}={\chi'\over 2}\equiv {\chi_0'\over 2}A^{-5/3}
\,;\quad {0.32\hbar\omega\over \Omega_p}\cong
{19.51\over A^{1/3}A_{mp}^{2/3}}= G \equiv {G_0}A^{-2/3}
\label{1s}
\end{equation}
and for two shells,
\begin{equation}
{0.216(2\hbar\omega)\over {\cal N}_p^2+{\cal N}_{p+1}^2}\cong
{1\over 2}\,{216\over A^{1/3}A_{c'p}^{4/3}}=
{\chi'\over 2}\equiv {\chi_0'\over 2}A^{-5/3}\,;\quad
{0.32(2\hbar\omega)\over \Omega_p+\Omega_{p+1}}\cong
{19.51\over A^{1/3}A_{c''p}^{2/3}}=G \equiv {G_0}A^{-2/3}
\label{2s}
\end{equation}
 where we have expressed the averaged norms in terms of $A_{c'p}$ and
$A_{c''p}$, both close enough to the the total number of
particles at the closure of shell $p$, $A_{cp}$, to be identifyed with it
in what follows.

For one shell the only problem comes from the conventional scalings, in
$A^{-5/3}$ and $A^{-2/3}$. It is possible to understand their origin: they
amount to set $A_{mp}=A$, which makes sense in comparing strengths in
distant regions, but is locally wrong. If taken at face value,
the $A^{-5/3}$ local behaviour predicts variations  in the moments of inertia
of neighbouring that are much larger than the observed ones
\cite{DU91,CHBWF92}.

In the two shell case the couplings  are reduced
with respect to the one shell values by a factor of about

\begin{equation}
(A_{mp}/A_{cp})^{k/3}=({2p+3 \over 2p+4})^k ,
\label{fafa}
\end{equation}

with $k=4$ for $\chi'$ and $k=2$ for$G$. If our identifications where correct
these discrepancies should not exist, and they are related to the risk we
described after eq. (\ref{ov}).
 There is no guarantee that we can
 approximate well the operator $\bar q_p+\bar q_{p+1}$ by
$ q_p+ q_{p+1}$. The compromise $\chi'$ in (\ref{2s}) is
too small for the lower shell but also too large for the upper one. If
the mixing is strong and both shells contribute equally to the
wavefunctions, then the large overlap in (\ref{ov}) indicates that the
compromise may work but in a nucleus well described by the lower shell
alone it makes no sense. The thing to do in this case is to
restrict the model to one shell.

\medskip

By overlapping the $q\cdot q$ form with a realistic interaction
Baranger and Kumar had obtained
$\chi'_0$= 203 {\it vs} our 216 for one shell, reduced by a factor of about
0.6 for two shells at $p=2$, close to the value of 7/8
 calculated in eq.(\ref{fafa}). When faced with this unwanted reduction,
instead of blaming the $q\cdot q$ form, they declared incorrect their method
of extraction  which - though in a primitive form - is identical to ours and
correct to within some details (discussed in section II). Then they
proceeded to invent another method to obtain the coupling constant,
unrelated to any interaction (see Note 2 at the end of the section).

At this point ref.\cite{BK68} becomes confusing because it is argued that
since {\em the} quadrupole force cannot be extracted from the realistic
interaction its origin must be something else {\it that may not
be quadrupole at all}.
What we are showing is that the ``something else'' is simply the normalized
quadrupole force.

 The extraordinary thing is that Baranger and Kumar had
found it! The reason they did not see what they had found is that they
reasoned in terms of the $A^{-5/3}$ scaling, in spite of having given the
correct argument about what the scaling must be!

\medskip

Now observe carefully eqs.(\ref{1s}) and (\ref{2s}) to discover the miracle
that repairs the dammage. The only sensible way to define $\chi'_0$ for one
shell is to ``equate'' $A_{mp}\approx A$, while for two shells we must take
 $A_{cp}\approx A$ and now $\chi'_0$ is identical to 216 in both cases!

{\em There is no contradiction between

 $\chi'$(1 shell)$\neq \chi'$(2 shells) at a given nucleus and

 $\chi_0'$(1 shell)$= \chi_0'$(2 shells), both calculated in different nuclei.}

\medskip

This is the subttle argument missing in \cite{BK68}
 It is only thanks to universal scaling that it can be made.

Everything seems to happen as if one flaw of the model
 - incorrect scaling - corrected the other: the space dependence of the
coupling constants. A better interpretation though, is that the model
should be restricted either to one shell in the vicinity of $A_{mp}$,
or to two shells in the vicinity of $A_{cp}$.

Now we note  that the rare earth
region studied in \cite{KUBA68} is approximately centered at the oscillator
closures Z=70 and N=112, which meets this restriction, and that the
 parameters used were (in Mev):
  \[\chi_0' \approx 280, \quad G_{0\pi}=27  \quad G_{0\nu}=22 ,\]
somewhat larger than those in eq.(\ref{2s}) but quite consistent with the
renormalized values of section III.
( The need to use different $G_0$ values
for neutrons and protons is readily explained by (\ref{2s}). It is a
mild manifestation of space dependence).

\medskip

We can draw two conclusions:

{\it i)} the conventional pairing plus quadrupole model in two shells
is far more consistent with the realistic interactions than its authors
believed.

{\it ii)} it can be made truly realistic by using the normalized operators, by
including a pair of other terms (octupole and hexadecapole), and by
examining more closely the monopole contribution. Very little of the basic
simplicity of the original will be lost in this improved version.

\medskip

{\em Note 1. On space dependence.}

Examine what happens in larger spaces when eq.(\ref{2s}) is generalized
to $M$ major shells.
 It is elementary to prove that $\chi'=O(A^{-1/3}M^{-4})$, and the
 overlap  tends to $\sqrt {5/9}$ and $\chi'=O(A^{-1/3}M^{-4})$. It means that
 to simulate - never mind how remotely - the behaviour of its realistic
 counterpart, the conventional $q\cdot q$ force must be affected by a
  vanishingly small $\chi'$.

\medskip

{\em Note 2. On the second method of extraction in \cite{BK68}.}
To replace direct extraction of $\chi'$ from the realistic interaction
Baranger and Kumar proposed a method  based on the idea that the average energy
of a nucleus is independent of its shape. It makes no reference to the force
nor to any empirical datum. It leads to:
 \[ \chi'={\hbar\omega\over {\cal N}_p^2+{\cal N}_{p+1}^2}\]
(eqs.(72,73) and paragraph following them in \cite{BK68}, but here  we have
introduced the norms instead of their asymptotic values).  This is
the same $\chi'$ of eq.(\ref{2s}), except that a factor $4\times 0.216$ has
been set to unity. The identity in form suggests that the method is a
valid dimensional analysis, but the similarity in the numbers is likely
to be accidental.

\begin{table}
\caption{Comparison of some matrix elements in the $sd$ shell
($5=d_{5/2},\;3=d_{3/2},\;1=s_{1/2}$) for different interactions.
$\leftarrow$ indicates large W-realistic discrepancies.}
\begin{tabular}{ccccccl}
\tableline
&rstu&JT&KLS&KB&Bonn$\;$B&W\\
\tableline
& 5553&10&3.03&3.17&3.31&2.54\\
&& 21&-0.52&-0.40&-0.22&-0.28\\
&& 30& 1.21& 1.87& 1.89& 2.22\\
&& 41&-1.24&-1.36&-1.28&-1.24\\
& 5533&01&-4.17&-3.79&-3.41&-3.19\\
&& 10& 1.45& 1.62& 1.29& 0.72\\
&& 21&-0.89&-0.90&-0.89&-1.62\\
&& 30& 0.13& 0.50& 0.56& 1.89$\;\;\leftarrow$\\
& 5153&20&-1.10&-1.44&-1.33&-0.1~$\;\leftarrow$\\
\tableline
\end{tabular}
\end{table}

\begin{table}
\caption{Lowest energies $E_\Gamma$ for $\Gamma=01$, 10, 20 and 30 and
$e^\gamma$ for all $\gamma$, and
overlaps of the wavefunctions for KLS, Bonn$\,$B and W.}
\begin{tabular}{ccccccccccccccc}
\tableline
&01&10&20&30&
&10&11&20&21&30&31&40&41\\
\tableline
&-5.42&-5.43&-2.68&-2.15&
 KLS &-2.18&~2.38&-2.90&-0.71&-0.82&~0.44&-1.61&~0.40\\
&-5.48&-6.24&-2.91&-2.66&
 Bonn\,B&-1.55&~2.64&-3.32&-0.97&-0.83&~0.46&-1.39&~0.52\\
&-5.69&-5.90&-0.95&-2.44&
 W&-2.16&~3.08&-3.18&-0.70&-0.94&~0.54&-1.60&~0.51\\
&~1.00&~1.00&~1.00&~1.00& $<$KLS$| $Bonn$\,$B$>$&
{}~~.99&~1.00&~1.00&~~.99&~~.98&~~.99&~1.00&~1.00\\
&~1.00&~0.98&~0.55&~0.82& $<$KLS$|$ W$>$&
{}~~.95&~~.99&~1.00&~~.98&~~.98&~~.92&~1.00&~1.00\\
\tableline
\end{tabular}
\end{table}

\newpage

\begin{figure}
  \begin{center}
    \leavevmode
    \psfig{file=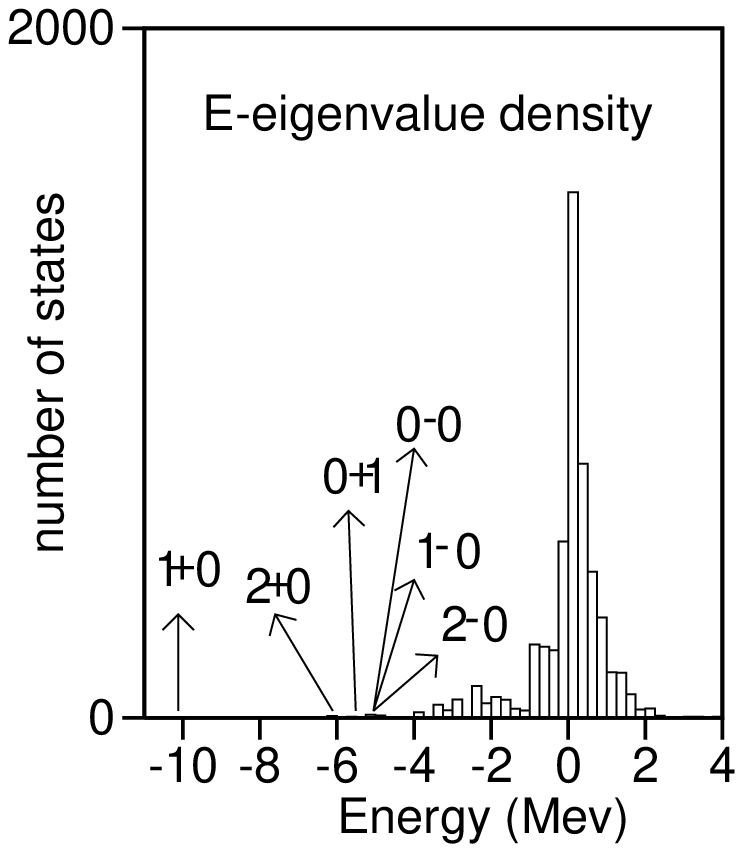,angle=270}
  \end{center}
\medskip
\caption{$E$-eigenvalue density for the KLS interaction in the pf+sdg
major shells $\hbar\omega=9$. Each eigenvalue has multiplicity $[\Gamma]$.
The largest ones are shown by arrows.}
\end{figure}

\begin{figure}
  \begin{center}
    \leavevmode
    \psfig{file=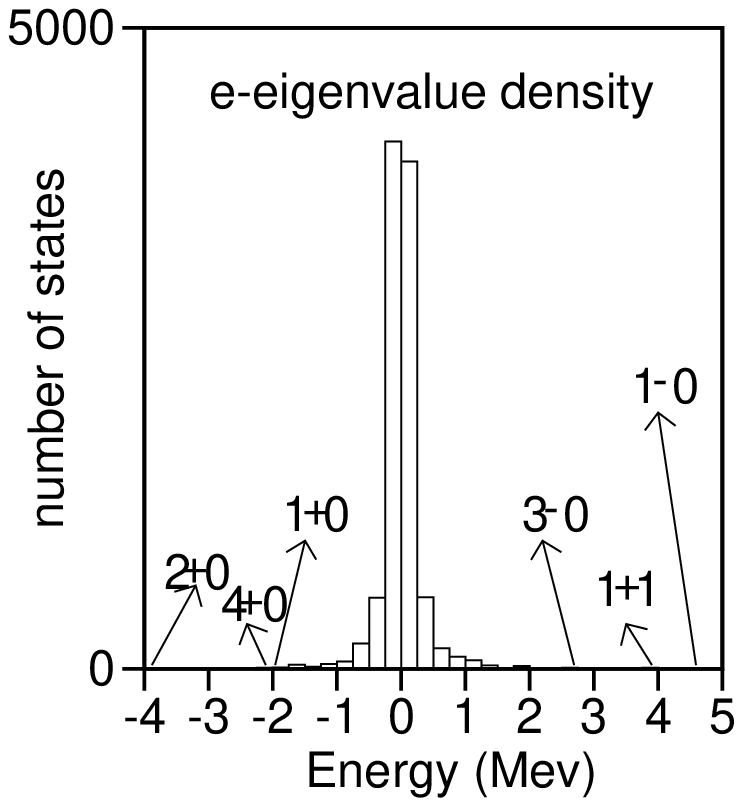,angle=270}
  \end{center}
\medskip
\caption{$e$-eigenvalue density for the KLS interaction in the pf+sdg
major shells. Each eigenvalue has multiplicity $[\gamma]$.
The largest ones are shown by arrows.}
\end{figure}

\newpage

\begin{figure}
  \begin{center}
    \leavevmode
    \psfig{file=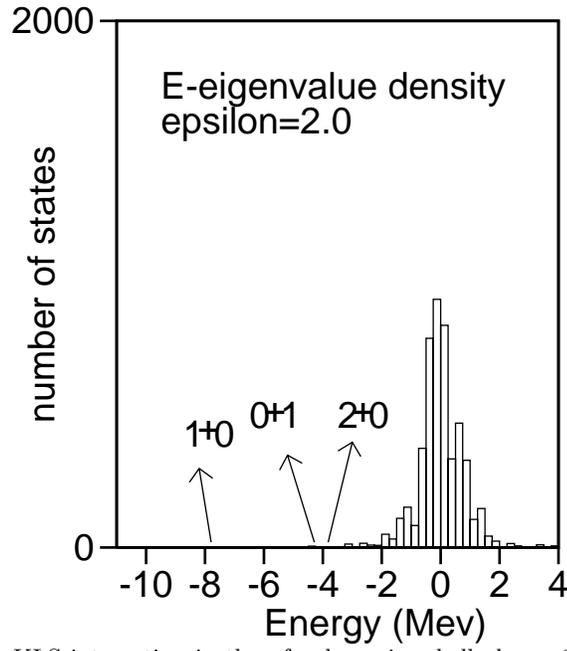,angle=270}
  \end{center}
\medskip
\caption{$E$-eigenvalue density for the KLS interaction in the pf+sdg
major shells $\hbar\omega=9$, after removal of the five largest multipole
contributions. Each eigenvalue has multiplicity $[\Gamma]$.
The largest ones are shown by arrows.}
\end{figure}

\begin{figure}
  \begin{center}
    \leavevmode
    \psfig{file=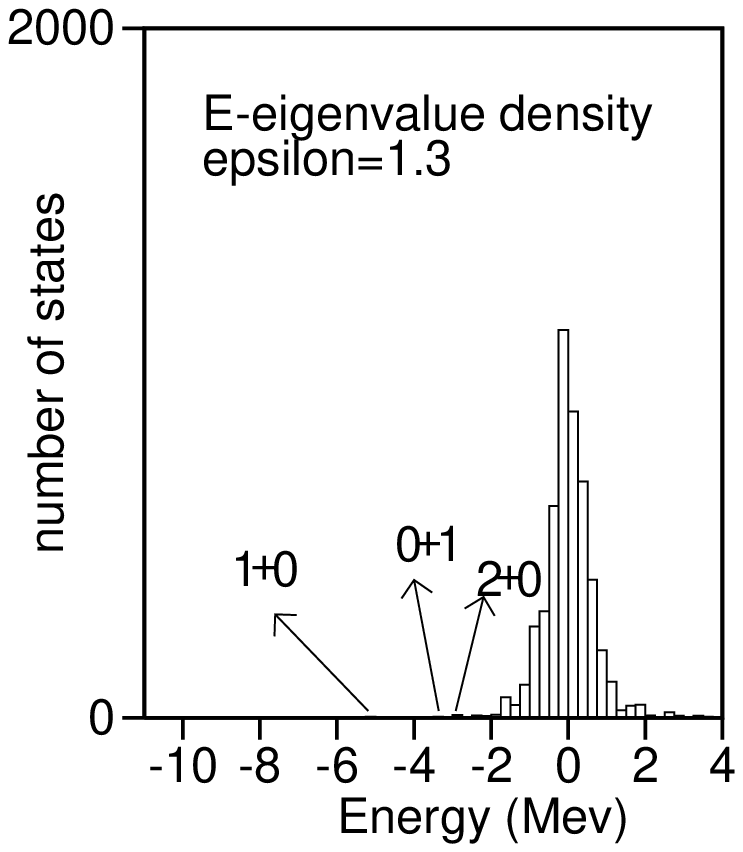,angle=270}
  \end{center}
\medskip
\caption{$E$-eigenvalue density for the KLS interaction in the pf+sdg
major shells $\hbar\omega=9$, after removal of multipole contributions
with $|e|>$~1.3. Each eigenvalue has multiplicity $[\Gamma]$.
The largest ones are shown by arrows.}
\end{figure}

\end{document}